\documentstyle[color,12pt,preprint,aps,tighten,epsfig,rotate,floats]{revtex}

\def\beq{\begin{equation}}
\def\eeq{\end{equation}}
\def\beqa{\begin{eqnarray}}
\def\eeqa{\end{eqnarray}}
\def\be{\begin{equation}}
\def\ee{\end{equation}}
\def\bea{\begin{eqnarray}}
\def\eea{\end{eqnarray}}

\begin{document}

\preprint{
\vbox{\hbox{IFIC-02-56}}}
\title{Improved Determination of the Electroweak Penguin Contribution 
to $\epsilon'/\epsilon$ in the Chiral Limit}
\author{Vincenzo Cirigliano}
\address{Departament de F\'isica Te\`orica, IFIC, Universitat de
Val\`encia - CSIC
\\
Apt. Correus 2085, E-46071 Val\`encia, Spain \\
Vincenzo.Cirigliano@ific.uv.es}
\author{John F. Donoghue and Eugene Golowich}
\address{Physics Department, University of Massachusetts\\
Amherst, MA 01003 USA \\
donoghue@physics.umass.edu,
golowich@physics.umass.edu \\}
\author{Kim Maltman}
\address{Department of Mathematics and Statistics, York University\\
4700 Keele St., Toronto ON M3J 1P3 Canada \\
and \\
CSSM, University of Adelaide \\
Adelaide, SA 5005 Australia \\
kmaltman@physics.adelaide.edu.au}
\maketitle
\thispagestyle{empty}
\setcounter{page}{0}
\begin{abstract}
\noindent We perform a finite energy sum rule analysis of the flavor
$ud$ two-point V-A current correlator, $\Delta\Pi (Q^2)$.  The analysis,
which is performed using both the ALEPH and OPAL databases for
the V-A spectral function, $\Delta\rho$, allows us to extract the 
dimension six V-A OPE coefficient, $a_6$, which is related to the matrix
element of the electroweak penguin operator, $Q_8$, by chiral symmetry.
The result for $a_6$ leads directly to the improved (chiral limit) 
determination $\epsilon^\prime /\epsilon = 
\left( - 15.0 \pm 2.7 \right) \cdot 10^{-4}$. Determination of
higher dimension OPE contributions also allows us to perform an
independent test using a low-scale constrained dispersive analysis,
which provides a highly nontrivial consistency check of the results.
\end{abstract}
\pacs{}

\vspace{1.0in}

\section{Introduction}
Intense effort carried out over many years 
to measure $\epsilon'/\epsilon$ has yielded 
the precise determination~\cite{nir} 
\beq 
\left[\epsilon'/\epsilon\right]_{\rm EXPT} = 
\left( 16.6 \pm 1.6 \right) \cdot 10^{-4} \ \ .
\label{expt}
\eeq
In the Standard Model, the primary dependence of $\epsilon'/\epsilon$ 
lies with the K-to-$2\pi$ matrix elements of 
the QCD penguin (${\cal Q}_6$) and EW penguin (${\cal Q}_8$) 
operators~\cite{buras1}.  For example, in the $\overline{MS}$-NDR 
scheme at scale $\mu =2$ GeV one has~\cite{ciuchini}  
\beqa
{\epsilon' \over \epsilon} &=& 20 \times 10^{-4}
\left( { {\cal I}m \lambda_t \over 1.3 \cdot 10^{-3}}\right)
\left[ -
2.0~{\rm GeV}^{-3} \cdot \langle (\pi\pi)_{I=0} |
{\cal Q}_6 | K^0\rangle_{2~{\rm GeV}} 
(1 - \Omega_{\rm IB}) \right. \nonumber \\
& & \left. \phantom{xxxxx} - 0.50~{\rm GeV}^{-3} \cdot 
\langle (\pi\pi)_{I=2} | {\cal Q}_8 | K^0\rangle_{2~{\rm GeV}} 
- 0.06 \right] \ \ ,
\label{r1}
\eeqa
where the factor $\Omega_{\rm IB}$ accounts for effects of 
isospin breaking. 
The electroweak penguin (EWP) operator, the subject of our attention  
in this paper, is given by 
\beq
{\cal Q}_8 \equiv {\bar s}_a\Gamma^\mu_{\rm L}d_b
\left( {\bar u}_b\Gamma_\mu^{\rm R}u_a - {1 \over 2}
{\bar d}_b\Gamma_\mu^{\rm R}d_a - {1 \over 2}
{\bar s}_b \Gamma_\mu^{\rm R}s_a\right) \ \ ,
\label{ewpop}
\eeq
where $\Gamma^{\mu}_{\rm L,R}  \equiv \gamma^{\mu} (1 \pm \gamma_{5})$
and $a,b$ are color indices.

In a previous paper~\cite{cdgm}, we worked 
in the SU(3) chiral limit ($m_u = m_d = m_s = 0$) 
and obtained for the EWP contribution to $\epsilon'/\epsilon$, 
\beq
\left[\epsilon'/\epsilon\right]_{\rm EWP} = 
\left( - 22 \pm 7 \right) \cdot 10^{-4} \ \ .
\label{old}
\eeq
The $32\%$ uncertainty in this determination is small enough 
to allow the conclusion that the chiral value for 
$\left[\epsilon'/\epsilon\right]_{\rm EWP}$
is negative and rather large (roughly the magnitude of the 
experimental signal for $\epsilon'/\epsilon$).  Moreover, we understand 
the source of the uncertainty.  Chiral sum rules used in Ref.~\cite{cdgm} 
to obtain the result in Eq.~(\ref{old}) require knowledge of V-A 
spectral functions over {\it all} energy.  However, data\footnote{In 
this paper we work with the ALEPH and OPAL spectral functions 
displayed in Figs.~1(a),1(b).  This normalization 
for $\Delta\rho$ corresponds to the flavor $ud$ two-point 
V-A current correlator and has {\it twice} the magnitude 
employed in Ref.~\cite{cdgm}.} provides input 
only up to the scale $s = m_\tau^2$.  We overcame this problem in 
Ref.~\cite{cdgm} by employing the chiral sum rules of
Weinberg~\cite{sw} and others~\cite{dgmly} as constraints.  In all, 
our procedure was subject to errors coming from the spectral 
function data set~\cite{aleph,opal} as well as 
those associated with quantities (the pion decay constant 
and the pion electromagnetic mass splitting) 
whose values must be estimated in the chiral limit.~\cite{abt} 

It is our purpose in this paper to report on an improved chiral 
determination of $\left[\epsilon'/\epsilon\right]_{\rm EWP}$
({\it i.e.} one with reduced uncertainty) which uses finite energy 
sum rules (FESR) as the main theoretical technique~\cite{vc2002}.  
However, it is necessary to summarize briefly aspects of Ref.~\cite{cdgm} 
since this new approach relies upon a number of details derived 
there.  We do this in Sect.~\ref{sect:summ}, then go on 
to describe the FESR analysis in Sect.~\ref{sect:fesr}, discuss 
the implications for the electroweak matrix elements 
in Sect.~\ref{sect:results}, and summarize our findings in 
Sect.~\ref{sect:concl}.
A more detailed discussion of the FESR analysis is presented in 
a companion paper~\cite{cdgm6}.

\vspace{-1.0cm}

\begin{picture}(50,150)(30,50)
\rotate[l]{
\psfig{figure=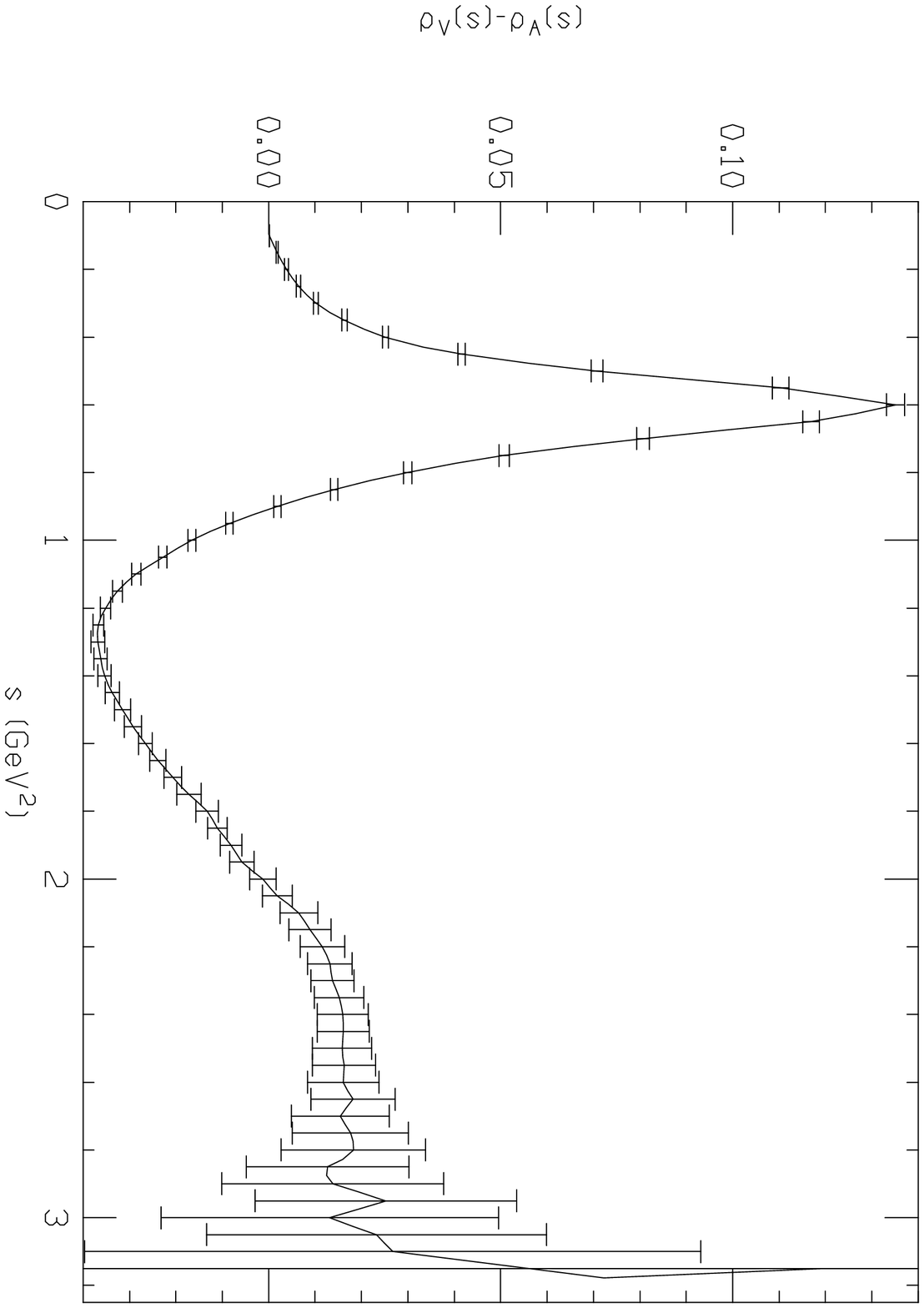,height=3.0in}
}
\hspace{1.0cm}
\rotate[l]{
\psfig{figure=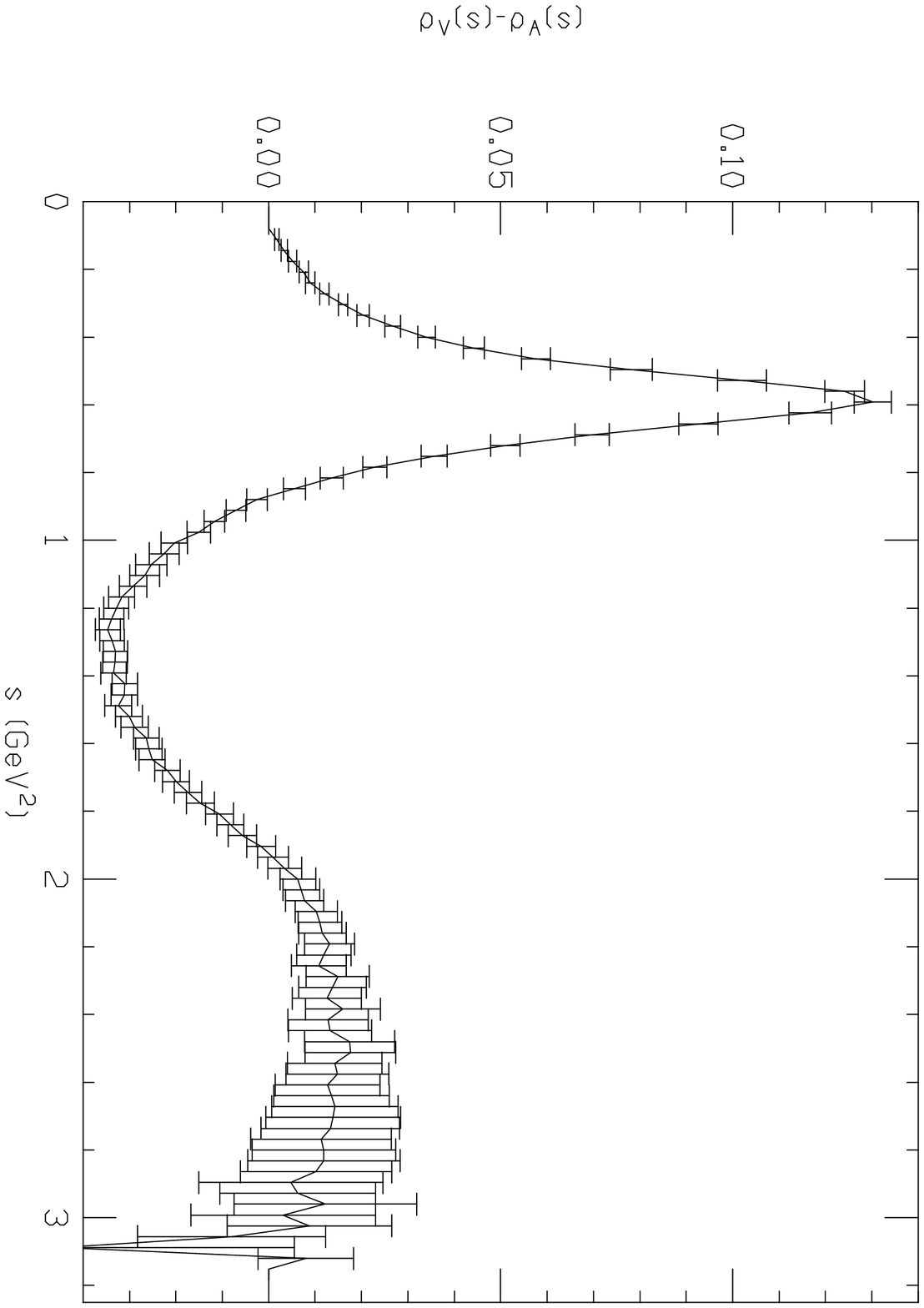,height=3.0in}
}
\put(-405,-15){{Fig.~1(a). ALEPH data}}
\put(-150,-15){{Fig.~1(b). OPAL data}}
\label{fig:data}
\end{picture}

\vspace{3.0cm}

\section{Chiral Properties of 
$\langle (\pi\pi)_{I=2} | {\cal Q}_{7,8} | K^0\rangle$}\label{sect:summ}

In the chiral limit the matrix elements 
$\langle (\pi\pi)_{I=2} | {\cal Q}_8 | K^0\rangle_\mu$ 
and also $\langle (\pi\pi)_{I=2} | {\cal Q}_7 | K^0\rangle_\mu$ 
become expressible in terms of vacuum matrix elements 
$\langle {\cal O}_1 \rangle_\mu$ and 
$\langle {\cal O}_8 \rangle_\mu$,~\cite{cdgm,dg}
\beqa
\lim_{p=0}~ \langle (\pi\pi)_{I=2} | {\cal Q}_7
|K^0\rangle_\mu & = & - {2 \over F_\pi^{(0)3}}~\langle {\cal O}_1
\rangle_\mu \ \ , \nonumber \\
\lim_{p=0}~ \langle (\pi\pi)_{I=2} | {\cal Q}_8 |
K^0\rangle_\mu & = & - {2 \over F_\pi^{(0)3}} ~\left[
{1 \over 3} \langle {\cal O}_1 \rangle_\mu +
{1 \over 2} \langle {\cal O}_8 \rangle_\mu \right]  \ \ ,
\label{r4}
\eeqa
where $F_\pi^{(0)}$ is the pion decay constant evaluated in the 
chiral limit.\footnote{The operators ${\cal O}_{1,8}$ are defined as 
\begin{eqnarray*}
{\cal O}_1 &\equiv& {\bar q} \gamma_\mu {\tau_3 \over 2} q
~{\bar q} \gamma^\mu {\tau_3 \over 2} q -
{\bar q} \gamma_\mu \gamma_5 {\tau_3 \over 2} q
~{\bar q} \gamma^\mu \gamma_5 {\tau_3 \over 2} q \ \ ,
\nonumber \\
{\cal O}_8 &\equiv& {\bar q} \gamma_\mu \lambda^a
{\tau_3 \over 2} q
~{\bar q} \gamma^\mu \lambda^a {\tau_3 \over 2} q -
{\bar q} \gamma_\mu \gamma_5 \lambda^a {\tau_3 \over 2} q
~{\bar q} \gamma^\mu \gamma_5 \lambda^a {\tau_3 \over 2} q
\ \ .
\end{eqnarray*}
In the above, $q = u,d,s$, $\tau_3$ is a Pauli (flavor) matrix,
$\{ \lambda^a \}$ are the Gell~Mann color matrices and the
subscripts on ${\cal O}_1$, ${\cal O}_8$ refer to
the color carried by their currents.}  The sum rule analysis of 
Ref.~\cite{cdgm} implies that the numerical effect 
of $\langle {\cal O}_1 \rangle_{2~{\rm GeV}}$ on 
$\lim_{p=0}~ \langle (\pi\pi)_{I=2} | {\cal Q}_8 |
K^0\rangle_{2~{\rm GeV}}$ is just $2.5\%$ that of 
$\langle {\cal O}_8 \rangle_{2~{\rm GeV}}$.  This means that 
at the level of accuracy we can realistically hope to achieve 
for $\lim_{p=0}~ \langle (\pi\pi)_{I=2} | {\cal Q}_8 
|K^0\rangle_{2~{\rm GeV}}$, it is sufficient to ignore the 
contribution from $\langle {\cal O}_1 \rangle_{2~{\rm GeV}}$ and 
focus on just the quantity $\langle {\cal O}_8 \rangle_{2~{\rm GeV}}$.
We return to this point at the end of this section.
\subsection{Getting $\langle {\cal O}_8 \rangle_\mu$ 
from the OPE of $\Delta \Pi$}\label{subsect:ope}

Much about $\langle {\cal O}_8\rangle_\mu$ (and also about $\langle 
{\cal O}_1\rangle_\mu$) can be learned from the flavor $ud$ V-A 
correlator. Defining $\Delta\Pi \equiv \Pi^{(0+1)}_{\rm V} - 
\Pi^{(0+1)}_{\rm A}$, where the superscript
$(0+1)$ indicates the sum of the spin $0$ and $1$ parts of the
relevant correlator, the dispersive representation of $\Delta\Pi$ is 
\beq
\Delta \Pi (Q^2)= - {2 F_\pi^2 \over m_\pi^2 + Q^2} + 
\int_{s_{\rm thr}}^\infty ds\ {\Delta \rho (s)  \over s + Q^2} 
\ \ .
\label{r7}
\eeq
Here, $Q^2 \equiv - q^2$ is the variable for space-like momenta and 
$\Delta \rho \equiv \rho_{\rm V}^{(0+1)}$$-$$\rho_{\rm A}^{(0+1)}$ 
is the spectral function of $\Delta\Pi$. 
For large space-like momenta $Q^2 \gg \Lambda_{\rm QCD}^2$ and 
through ${\cal O} (\alpha_s^2)$ in QCD counting, 
$\Delta \Pi (Q^2)$ can be represented via the operator product 
expansion (OPE) 
\beq
\Delta \Pi (Q^2) \sim \displaystyle\sum_d \ {1 \over Q^d} \left[
a_d (\mu) + b_d (\mu) \ln{Q^2 \over \mu^2} \right] \qquad
(d = 2,4,6,8,10, \dots) \ \ ,
\label{r11}
\eeq
where $a_d (\mu)$ and $b_d (\mu)$ are combinations of vacuum 
expectation values of local operators of dimension $d$.  
In the chiral limit, the contributions from dimensions $d=2,4$ are 
absent and the above sum begins with dimension $d=6$. 

The $d=6$ OPE coefficients $a_6, b_6$ are of special interest 
because they are related to the vacuum matrix elements 
$\langle {\cal O}_8 \rangle_\mu$ and $\langle {\cal O}_1 \rangle_\mu$. 
Since the relations 
are renormalization scheme dependent we consider for definiteness 
${\rm {\overline {MS}}}$ renormalization with NDR and HV 
prescriptions for $\gamma_5$ and the evanescent operator basis used 
in Refs.~\cite{buras2,martinelli}. Including ${\cal O} (\alpha_s^2)$ 
terms this leads to 
\beqa
a_6 (\mu) &=& 2\, \left[ 2 \pi \langle \alpha_s {\cal O}_8 \rangle_\mu +
 A_8 \langle \alpha_s^2 {\cal O}_8 \rangle_\mu +
A_1 \langle \alpha_s^2 {\cal O}_1 \rangle_\mu \right] \ \ ,\nonumber \\
b_6 (\mu) &=& 2\, \left[ B_8 \langle \alpha_s^2 {\cal O}_8 \rangle_\mu +
B_1 \langle \alpha_s^2 {\cal O}_1 \rangle_\mu \right] \ \ .
\label{r15a}
\eeqa
In terms of their dependence on the NDR or HV renormalization 
scheme, the coefficients $A_1,A_8$ and $B_1,B_8$ are given by 
\begin{eqnarray}
\begin{array}{c||c|c}
   & {\rm NDR} & {\rm HV} \\ \hline
A_1 & 2 & -10/3 \\
A_8 & 205/36 & 169/36 \\ 
B_1 & 8/3 & 8/3 \\
B_8 & -2/3 & -2/3 \\ 
\end{array}
\label{scheme0} 
\end{eqnarray}
where we work with three colors ($N_c = 3$) and four active flavors 
($n_f = 4$). 

Given our earlier discussion following Eq.~(\ref{r4}) about 
the dominance of $\langle {\cal O}_8 \rangle_{2~{\rm GeV}}$ 
relative to $\langle {\cal O}_1 \rangle_{2~{\rm GeV}}$,  
it follows from Eq.~(\ref{r15a}) 
that an attractive alternative to the sum rule procedure of 
Ref.~\cite{cdgm} for finding $\lim_{p=0}~ \langle (\pi\pi)_{I=2} 
| {\cal Q}_8 |K^0\rangle_\mu$ is to 
determine  the OPE coefficient $a_6(\mu)$.
The FESR approach is naturally structured to do just that.  
Moreover, since only knowledge of $\Delta \rho$ in the data region 
is required, the prospect of an improved determination is well 
motivated.
\subsection{The Sum Rule Approach Revisited}\label{subsect:sumrule}
Independently from the method described in the preceding section, 
knowledge gained from the FESR analysis of 
the higher dimensional OPE coefficients $\{ a_{d\ge 8} \}$ 
allows us to return to the sum rule method of Ref.~\cite{cdgm} 
and perform an improved determination 
of $\langle {\cal O}_8 \rangle_\mu$ and $\langle
{\cal O}_1 \rangle_\mu$.  This hybrid method provides a nontrivial 
consistency check on the FESR results of this paper.  
We summarize this approach in the following.  

The two sum rules derived in Ref.~\cite{cdgm} are 
\beqa 
& & \langle {\cal O}_1 \rangle_\mu - {3 C_8 \over 8 \pi}  
\langle \alpha_s {\cal O}_8 \rangle_\mu = {3 \over (4 \pi)^2 } 
\left[ I_1 (\mu) + H_1 (\mu) \right] \nonumber \\
& & 2 \pi \langle \alpha_s {\cal O}_8 \rangle_\mu 
+ A_1 \langle \alpha_s^2 {\cal O}_1 \rangle_\mu +  
A_8 \langle \alpha_s^2 {\cal O}_8 \rangle_\mu = 
I_8 (\mu) - H_8 (\mu) \ \ , 
\label{sumrule}
\eeqa
where
\beq
I_1 = {1 \over 2} \int_0^\infty ds\ s^2 \ln \left({s + 
\mu^2 \over s} \right) ~\Delta\rho (s) \ , \qquad 
I_8 = {1 \over 2} \int_0^\infty ds\ s^2 {\mu^2 \over s + \mu^2}
~\Delta\rho (s) 
\label{integral}
\eeq
and 
\beq
H_1 = {1 \over 2} \sum_{d \ge 8} {2 \over d - 6}\cdot 
{a_d (\mu) \over \mu^{d - 6}} \ , \qquad 
H_8 = {1 \over 2} \sum_{d \ge 8} 
{a_d (\mu) \over \mu^{d - 6}} \ \ .
\label{higherope}
\eeq
The above definitions of $I_{1,8}$ and $H_{1,8}$ coincide with the 
ones given in Ref.~\cite{cdgm} --- the prefactors of 1/2 simply account 
for the different normalization of $\Delta \rho$ used in this work. 
The scheme dependent quantities $A_1, A_8$ of 
Eq.~(\ref{sumrule}) are given in Eq.~(\ref{scheme0}) and $C_8$ 
appears in Table~1 of Ref.~\cite{cdgm}.  Observe that even if 
the higher dimension OPE coefficients in $H_{1,8}$ are not known, 
the sum rules can nonetheless be successfully evaluated.  
One simply chooses a sufficiently large scale $\mu$, 
{\it e.g.} at $\mu = 4$~GeV, to suppress contributions from the 
$\{ a_{d\ge 8} \}$ and then uses the renormalization group equations 
to evolve down to the scale $\mu = 2$~GeV.  This was the procedure 
adopted in Ref.~\cite{cdgm}.  However, it turns out that the 
uncertainty in evaluating the spectral integrals in Eq.~(\ref{sumrule}) 
{\it grows} with increasing $\mu$ and this source of error is 
ultimately communicated to the sum rule prediction for 
$\langle (\pi\pi)_{I=2} | {\cal Q}_{7,8} | 
K^0\rangle^{\overline{\rm MS}}_{\mu=2 {\rm GeV}}$.  
Our FESR determination of the $\{ a_{d\ge 8} \}$ allows us to 
avoid this difficulty by allowing evaluation of the sum rules 
directly at $\mu = 2$~GeV and thus leads to reduced uncertainties 
in the matrix element values.  Results of this hybrid approach 
are presented in Sect.~\ref{sect:results}. 

\vspace{0.3cm}

We now turn to a discussion of our FESR analysis.  

\section{FESR Analysis}\label{sect:fesr}
We begin with the following exact consequence of Cauchy's 
theorem (plus the rigorous statement of analytic structure 
for $\Delta \Pi$), 
\beq
\int_{s_{\rm th}}^{s_0} ds \ w(s) ~ \Delta \rho (s) 
- 2 F_\pi^2 w(m_\pi^2) = 
-{1 \over 2 \pi i} \oint_{|s|=s_0} 
ds \ w(s)~ \Delta \Pi (s) \ \ ,\label{du1}
\eeq
where the contour of integration for $\Delta \Pi$ 
is a circle of radius $s_0$ ({\it cf.} Fig.~2) 
and $w(s)$ is analytic on and within the given contour. 
The additive term proportional to $F_\pi^2$ on the left hand 
side of Eq.~(\ref{du1}) arises from the pion pole.
The object of FESRs is to replace $\Delta \Pi$ on the circle 
by the asymptotic OPE form $\Delta \Pi_{\rm OPE}$ (see Eq.~(\ref{r11})) 
and thereby obtain constraints on the OPE coefficients in terms of data 
from knowledge of the spectral function $\Delta \rho$.  

\vspace{1.0cm}

\begin{picture}(50,150)(-130,0)
\psfig{figure=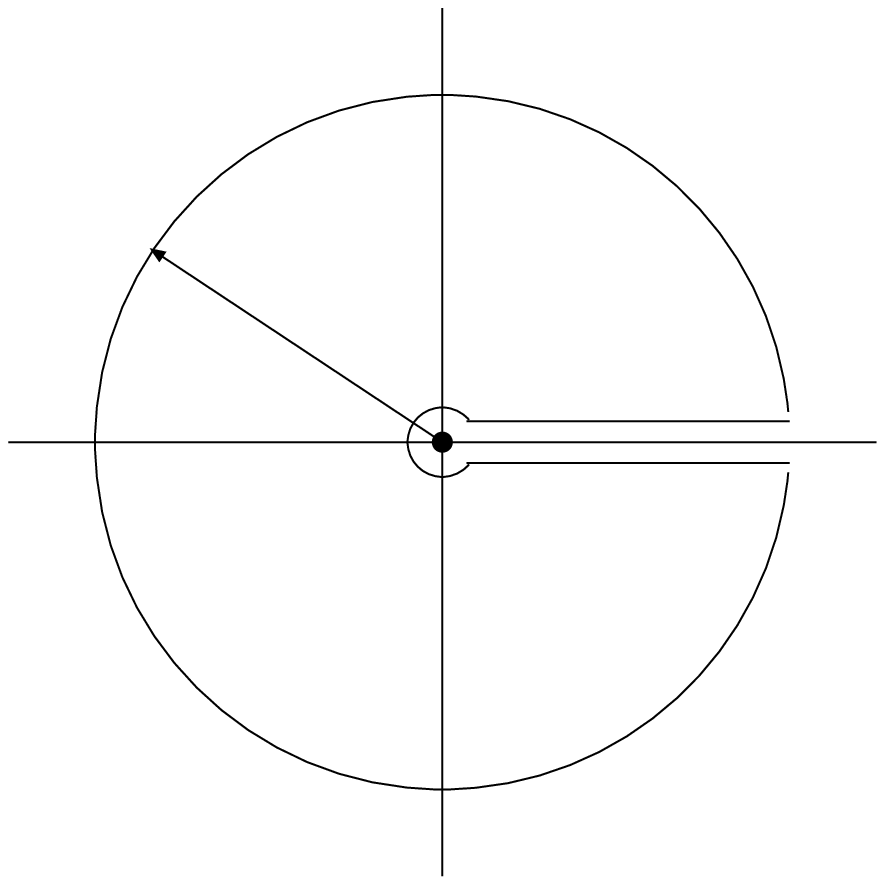,height=2.5in}
\put(-150,-15){{Fig.~2. FESR contour.}}
\label{fig:fesr}
\end{picture}

\vspace{1.0cm}

Implementing this procedure but making no additional assumptions, 
we can then write compactly 
\beq
\int_{s_{\rm th}}^{s_0} ds \ w(s) ~ \Delta \rho (s) 
- 2 F_\pi^2 w(m_\pi^2) = 
-{1 \over 2 \pi i} \oint_{|s|=s_0} 
ds \ w(s)~ \Delta \Pi_{\rm OPE} (s) - R [s_0,w] \ \ , 
\label{du3} 
\eeq
where the quantity $R [s_0,w]$ is defined as  
\beq 
R [s_0,w] \equiv 
-{1 \over 2 \pi i} \oint_{|s|=s_0} 
ds \ w(s)~ \left[ \Delta \Pi_{\rm OPE} (s) 
- \Delta \Pi (s) \right] \ \ . 
\label{du4}
\eeq
While Eqs.~(\ref{du3}),(\ref{du4}) are valid for any weight 
$w(s)$ analytic in a region including the contour of Fig.~2, 
the presence of the {\it a priori} unknown additive term 
$R [s_0,w]$ generates a systematic uncertainty in the 
extraction of the OPE coefficients $a_d$ via Eq.~(\ref{du3}). 
Therefore it is highly desirable to work with a
range of $s_0$ values and with weight functions $w(s)$ such that the
remainder term $R [s_0,w]$ is small compared to the spectral integral
itself. 
To accomplish this, we rely on the observation that 
for $|s|$ large enough ($s_0 \gg \Lambda_{QCD}^2$), the OPE 
provides a good representation of the full correlator along the whole
circle except in a region localized around the time-like axis.  The
physics of this breakdown is given by the arguments of Poggio, Quinn
and Weinberg~\cite{poggio}.
As a consequence one expects that weights with a zero at $s=s_0$, 
de-emphasizing the region where the OPE fails, are good candidates 
to generate a small-sized $R [s_0,w]$ even for the relatively low 
$s_0$ values, between $2$ GeV$^2$ and $m_\tau^2$, as used in 
our FESR analysis.  Arguments which justify the choice of this range 
are given in Sect.~\ref{subsect:fit} following Eq.~(\ref{chisq}).

Supporting evidence for the procedure outlined above is provided 
by one of us in previous studies of the V and A 
correlators~\cite{km,vc2002}.  Internal 
consistency checks of the V-A analysis are also possible (see Section 
IIIB for a further discussion of this point). Moreover, a study of 
models with pole contributions which violate 
duality shows that the method succeeds in extracting the 
asymptotic values of the corresponding $a_d$ coefficients, 
even in the presence of such duality violation.~\cite{cdgm6}

\subsection{Choice of FESR Weights}\label{subsect:weight}
To complete preparations for performing our FESR study, 
the highly nontrivial procedure of constructing appropriate 
weights $w(s)$ must be addressed.  Three basic considerations
govern our choice of weights:
\begin{enumerate}
\item Because the OPE is known to fail along the ${\cal R}e \, s \ge 0$ 
axis~\cite{poggio}, we employ only `pinched' weights $w(s_0) = 0$. 
In particular, recall that the kinematic weight 
($w(y_\tau )=(1-y_\tau )^2(1+2y_\tau )$ with $y_\tau \equiv
s/m_\tau^2$), which must be unfolded from the experimental decay 
distribution in order to obtain $\Delta\rho$, 
contains a double zero at $s=m_\tau^2$.   
To avoid enhancements of contributions from the endpoint 
region, where experimental errors are large, we therefore employ 
weights $w(y) = (1-y)^2 p(y)$ ($y \equiv s/s_0$ and $p(y)$ is a 
polynomial) which also have a double zero at $s=s_0$.
An earlier study~\cite{km} has demonstrated that the suppression 
thus produced suffices to circumvent the OPE breakdown.
\item In order to maximize the statistical signal, the weights should
be such that large cancellations in the difference between the separate 
V and A spectral integrals are avoided.
\item The separation of contributions with different dimension 
in a given pinch-weighted FESR (pFESR) relies on the fact that 
the integrated OPE contributions with different $d$ scale differently
with $s_0$.  When one works in a limited window of 
$s_0$ values, the accuracy with which one can perform this separation
decreases as the number of contributing $a_d$ terms increases.
We, therefore, work with weights such that the minimum number (two) of
$a_d$ contributions occurs in any of our sum rules.
\end{enumerate}
The above points are elaborated upon in Ref.~\cite{cdgm6}.

The FESR equations for a set of weights, $\{ w_n\}$, are 
of the general form 
\beq
J_n (s_0) = f_n (a_d, a_{d'}, b_d, b_{d'}; s_0)
 \ \ .
\label{fesr}
\eeq
The quantity $J_n (s_0)$ is defined as 
\beq
J_n (s_0) \equiv  \int_0^{s_0} ds \ w_n(y) \Delta \rho(s) - 
2 F_\pi^2 w_n(m_\pi^2) - c_n^{(4)}  \ \ , 
\label{jn}
\eeq
where $y \equiv s/s_0$ and $c_n^{(4)}$ is the contribution 
from the dimension-four part of the OPE.\footnote{All the 
$c_n^{(4)}$ terms ($n=1,2$) are known explicitly up to 
$O (\alpha_s^2)$ and are numerically small, being 
proportional to the pion mass~\cite{cdgm6}. The  $2 F_\pi^2 
w_n(m_\pi^2)$ terms are likewise small in our analysis, except 
for the case of the weight $w_1$, given below in Eq.~(\ref{w1}).}

The functions $f_n$ represent the OPE terms with $d\ge 6$ 
integrated on the $|s|=s_0$ circle.
Polynomial weights of increasing degree allow one to extract
condensates $a_d$ of increasing dimension. We use as starting point
for our analysis the sum rules based on the weights of lowest degree
satisfying the above criteria. We have also explored generalizations to 
higher degree polynomials, and report details of this analysis  
elsewhere~\cite{cdgm6}. 

The lowest degree weights satisfying our criteria turn out 
to have degree three.  The one which produces the smallest relative 
errors on the spectral integrals is 
\beq
w_1 (y) = (1 - y)^2  \, (1 - 3 y) \ \ .
\label{w1}
\eeq
An independent choice, also giving small relative errors, is 
\beq
w_2 (y) = (1 - y)^2  \, y  \ \ .
\label{w2}
\eeq
These weights produce OPE integrals involving $a_6$ and $a_8$:
\beqa
& & f_1(a_6,a_8;s_0) \equiv 
{7 \over s_0^2} ~ a_6 ~ 
\left[ 1 + r_6 \, \log \left(
{s_0 \over \mu^2} \right) \ + {3 \over 14} \, r_6  \right]
\ + \ 3 \,  {a_8\over s_0^3} \nonumber \\
& & f_2(a_6,a_8;s_0) \equiv -  {2 \over s_0^2}~a_6 
 \, \left[ 1 + r_6 \, \log \left(
{s_0 \over \mu^2} \right) \right] \  - \ {a_8\over s_0^3} \ \  . 
\label{a6a8}
\eeqa
In Eqs.~(\ref{a6a8}) $r_6 \equiv b_6/a_6$, and we neglect all the
other $\{ b_{n} \}$ OPE coefficients since they are QCD suppressed
($b_n \sim a_n \alpha_s/\pi$), and are accompanied by small numerical
coefficients in the FESR relations.  For the ratio $r_6$ we rely 
on the explicit NLO QCD expressions of Eq.~(\ref{r15a}) and on 
the numerical estimates of $\langle {\cal O}_1\rangle$
and $\langle {\cal O}_8\rangle$ given in Ref.~\cite{cdgm}.

\subsection{FESR Analysis: Fit and Results}\label{subsect:fit}

The FESR relations of Eqs.~(\ref{fesr}),(\ref{jn}),(\ref{a6a8}) 
for the cases $n=1,2$ 
allow us to impose constraints on the coefficients $a_6$ and 
$a_8$, as we know $J_{1,2} (s_0)$ 
from data for $s_0 \leq m_\tau^2$. 
The treatment of Eq.~(\ref{jn}) requires some care due to the
presence of strongly correlated errors in the experimental spectral
function $\Delta\rho$.  
Using the covariance matrix for the spectral function 
data, we calculate the covariance matrix for the 
set of spectral integrals $J_n (s_0)$: 
$\mbox{Cov} ( J_n (s_0), J_m (s_0 ') ) \equiv 
{\bf V}^{(n,m)}_{s_0,s_0'} $. 
Thus we form the weighted least-square function 
\beq
\chi^2 = \displaystyle\sum_{s_0}\sum_{n=1,2} 
\left( J_n (s_0) - f_n (s_0) \right)
\left[{\bf V}^{(n,n)}_{s_0,s_0}\right]^{-1} 
\left( J_n (s_0) - f_n (s_0) \right) \ \ , 
\label{chisq}
\eeq
which sums over the set of $s_0$ values and the 
two FESR relations ($n = 1,2$). 

We determine the `window' of $s_0$ values used in our analysis 
as follows, considering for definiteness the ALEPH data sample.
The largest $s_0$ value (upper edge of 
the window) is taken as $s_0 = 3.15~{\rm GeV}^2 \simeq m_\tau^2$ 
simply because it is there that the ALEPH data sample for $\Delta\rho$ 
runs out.  The smallest $s_0$ value, taken as $s_0 = 
1.95~{\rm GeV}^2$, is established 
by trying ever lower $s_0$ values until the extracted 
$a_6 (s_0)$ (which is the most accurately determined OPE coefficient) 
ceases to be consistent with the previous values obtained from smaller 
analysis windows.  In the final analysis seven equally-spaced values, 
from  $s_0 = 1.95$ to $s_0 = 3.15$ in the ALEPH case, 
were adopted~\cite{cdgm6}. 
We use the `diagonal' least-square function as defined in
Eq.~(\ref{chisq}) to avoid known problems of fits to strongly
correlated data~\cite{dagostini}.  Minimization of $\chi^2$ with
respect to variations in $a_6$ and $a_8$ yields best fit values for
these quantities. 
With this procedure, as is well known, the one-sigma errors and 
rms errors do not coincide. The former are smaller, and underestimate 
the variation of the fitted $a_d$  produced by fluctuations in the 
input experimental data. All the errors quoted below are the rms errors, 
{\it i.e.} the square roots of the diagonal elements of the covariance 
matrix for the ${a_6,a_8}$ solution set. 
We thereby arrive at the following results with {\it renormalization 
scale set at $\mu= 2 \,{\rm GeV}$}:

\vspace{0.2cm}

\noindent {\bf 	Fit to ALEPH data}: 
\beq
a_6 =  \left( - 44.5 \pm 6.3 \pm 3.4 \right) \cdot 10^{-4}~{\rm GeV}^6 
\qquad 
a_8 =  \left( - 61.6 \pm 28.9 \pm 13.8 \right) 
\cdot 10^{-4}~{\rm GeV}^8 \ .
\label{a6a8al}
\eeq
The first error is associated with the ALEPH covariance matrix.  
The second is obtained by adding in quadrature uncertainties from 
parameters which enter the ALEPH normalization of the spectral 
function and from parameters occurring on 
the `theoretical' side of the FESR.   
The correlation coefficient for the fitted parameters is found to be 
$c (a_6, a_8) = -0.995$, so the output is highly correlated.

By extending the set of weights employed in our analysis, it is possible 
to construct sum rules which receive contributions from the $a_d$
with $d>8$.\footnote{The process of choosing weights in such a way as to 
optimize the determination of the higher dimension condensate
combinations, $a_d$, is discussed in more detail in
Ref.~\cite{cdgm6}.}  In this manner we have successfully determined 
the $a_d$ up to $d=16$ and thus have evaluated the quantities 
$H_1$ and $H_8$ defined earlier in Eq.~(\ref{higherope}).  We find 
at the renormalization scale $\mu =2$~GeV, 
\beq
H_1 = \left( - 2.7 \pm 3.6 \pm 0.5 \right) \cdot 10^{-4}~{\rm GeV}^6 
\qquad 
H_8 =  \left( - 1.3 \pm 3.0 \pm 2.0 \right) 
\cdot 10^{-4}~{\rm GeV}^6 \ .
\label{h1h8al}
\eeq
The first error in $H_{1,8}$ comes from the correlated 
uncertainties in the OPE coefficients $\{ a_d \}$.  The second 
represents our estimate of the error incurred in truncating the 
OPE sum at $d=16$.  We have obtained this estimate by extending 
our extraction out to $d=24$.  Although the sums for 
$H_{1,8}(2~{\rm GeV})$ are found to be well converged by $d=24$, 
the extractions of the $a_d$ with $d=18 \to 24$ are, however, less 
certain. We therefore use the difference of the $d=16$ and $d=24$ 
sums as a means of estimating the error associated with employing 
only the better-determined $d \le 16$ terms.~\cite{cdgm6}

\vspace{0.2cm}

\noindent {\bf Fit to OPAL data}: 
\beq
a_6 = \left( - 54.3 \pm 7.2 \pm 3.1 \right) \cdot 10^{-4}~{\rm GeV}^6 
\quad 
a_8 =  \left( - 13.5 \pm 33.3 \pm 11.7 \right) 
\cdot 10^{-4}~{\rm GeV}^8 ~.
\label{a6a8op}
\eeq
The first error is associated with the OPAL covariance matrix, 
and the second is obtained by adding in quadrature the uncertainties
associated with parameters entering both the normalization of the
spectral function as well as those from the `theoretical' side of the FESR. 
The correlation coefficient for the fitted parameters is found to be 
$c (a_6, a_8) = -0.989$, again corresponding to a highly correlated 
output. 

In like manner to the ALEPH-based determinations of $H_1$ and $H_8$, 
we have results from OPAL data:
\beq
H_1 = \left(- 0.4 \pm 3.7 \pm 0.1 \right) \cdot 10^{-4}~{\rm GeV}^6 
\qquad 
H_8 =  \left( 0.6 \pm 3.0 \pm 1.0 \right) 
\cdot 10^{-4}~{\rm GeV}^6 \ .
\label{h1h8op}
\eeq
The errors are treated analogously as in Eq.~(\ref{h1h8al}).

\vspace{0.5cm}

Upon adding in quadrature the uncertainties displayed above, 
we obtain the determinations gathered in Table~1.
\begin{eqnarray*}
\begin{array}{c||c|c}
    \multicolumn{3}{c}{\rm Table~ 1:\ Summary\ of\ FESR\ Results} 
\\ \hline
   & a_6 ({\rm GeV}^6) & a_8 ({\rm GeV}^8) \\ \hline
{\rm ALEPH} & (-44.5 \pm 7.2 ) \cdot 10^{-4} 
& (-61.6 \pm 32.0 ) \cdot 10^{-4} \\
{\rm OPAL} & (-54.3 \pm 7.8 ) \cdot 10^{-4} 
& (-13.5 \pm 35.3) \cdot 10^{-4} \\
\end{array}
\end{eqnarray*}

\subsection{Testing for Duality Violation}\label{subsect:duality}

The errors reported above are essentially of an experimental nature.
In order to address potential systematic effects due to $R [s_0,w]$
(duality violation), we have repeated the fit procedure in different
$s_0$ windows (nested sub-windows and non-overlapping sub-windows). We
find that the new values for the fitted parameters are very consistent
with each other, thereby confirming that the effect of $R [s_0,w]$ is
suppressed in this case. 
A more explicit and revealing portrait of the FESR machinery is
obtained by plotting $J_{n} (s_0)$ and $f_n (a_6, a_8; s_0)$ as a
function of $s_0$, as in Figs.~3-4. In each case, an excellent match
of the OPE curve versus data is achieved, showing no sign of duality
violation within the present experimental uncertainty.

\begin{picture}(50,130)(30,50)
\psfig{figure=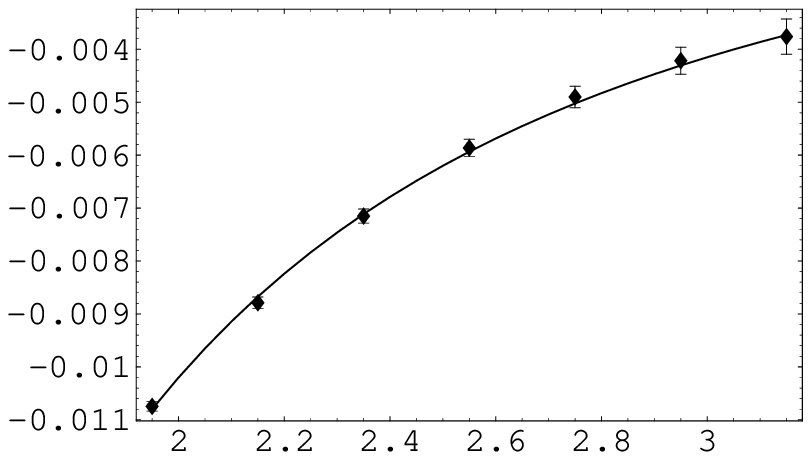,height=2.0in}
\psfig{figure=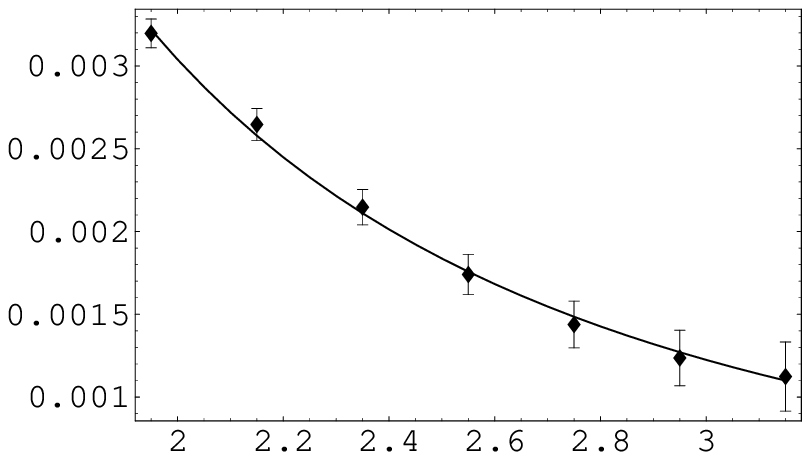,height=2.0in}
\put(-450,-15){{Fig.~3. Variation of $f_1$ (continuous curve)}}
\put(-410,-30){{and $J_1$ (data) with $s_0~({\rm GeV}^2)$.}}
\put(-210,-15){{Fig.~4. Variation of $f_2$ (continuous curve)}}
\put(-170,-30){{and $J_2$ (data) with $s_0~({\rm GeV}^2)$.}}
\end{picture}
\vspace{3.6cm}

Given the relatively low values of $s_0$ used in the analysis, a
legitimate worry is that the FESR analysis, despite the excellent
match displayed in Figs.~3-4, might just be extracting the
coefficients relevant to the Laurant expansion of the correlator in a
sub-asymptotic regime, and not the condensates relevant to the
truly asymptotic region. 
A way to test whether or not we are in such a scenario is to use the
condensates extracted in the FESR analysis as input in a dispersive
analysis relevant to the asymptotic regime, and observe the 
quality of the match there.   Such a class of tests is readily
possible within explicit models~\cite{golper,beane} of a given
correlator, and shows a very mild pattern of duality violation. 
Any attempt to perform such tests in the real world 
is limited by incomplete  knowledge of the spectral function. 
Nonetheless, in our case, with the help of classical sum rules 
it is possible to construct relevant asymptotic tests,  
which are described in Ref.~\cite{cdgm6}. 
All indications coming from these tests point to negligible 
duality violation, within the present errors. 

\section{The Electroweak Matrix Elements
$\langle (\pi\pi)_{I=2} | {\cal Q}_{7,8} | K^0\rangle$}\label{sect:results}
In the previous section, we have used the FESR machinary to obtain 
numerical determinations of OPE coefficients through order $d=16$.  
These can be used to obtain the matrix elements ${\cal M}_{7,8}$, 
\beq
{\cal M}_{7,8} \equiv { \langle (\pi\pi)_{I=2} | {\cal Q}_{7,8} 
| K^0\rangle_{\mu = 2~{\rm GeV}}^{\overline{\rm MS}-{\rm NDR}}
\over 1~{\rm GeV}^3} \ \ .
\label{matele}
\eeq
Note in this definition that ${\cal M}_{7,8}$ are simply dimensionless 
numbers.  We apply our FESR results via two distinct procedures: 
\begin{enumerate}
\item FESR: Insertion of $a_6$ into Eqs.~(\ref{r4}),(\ref{r15a}) 
leads directly to ${\cal M}_{8}$, up to a small contribution 
(at the $5\% $ level) due to  $\langle O_1 \rangle$.
In the numerical evaluation we use $\langle O_1 \rangle$ 
from Ref.~\cite{cdgm}. 

\item HYBRID: Insertion of the FESR-derived quantities $H_{1,8}$ in the 
sum rules of Eq.~(\ref{sumrule}), together with Eq.~(\ref{r4}), 
leads to both ${\cal M}_{7}$ and  ${\cal M}_{8}$. 
This `hybrid' approach requires as input the integrals 
\beq
I_1 = - (39.7 \pm 3.1) \cdot 10^{-4}~{\rm GeV}^6\ ,  \qquad 
I_8 = - (26.2 \pm 3.0) \cdot 10^{-4}~{\rm GeV}^6 \ \ , 
\label{i1i8}
\eeq
each evaluated at scale $\mu = 2$~GeV ~\cite{cdgm}.
\end{enumerate}

In Eqs.~(\ref{r15a}),(\ref{sumrule}) we use the $n_f=4$
matching coefficients, appropriate for the renormalization 
scale $\mu = 2$ GeV.  Moreover, in relating ${\cal M}_{7,8}$ 
to $\langle O_{1,8} \rangle $  (see Eq.~(\ref{r4})) we 
use $F_{\pi}^{(0)} = 87.2 \pm 2.6 $ MeV. 
We collect our results in Table~2, and for the sake of completeness 
include our previous sum rule determination (the `RWM' approach of 
Ref.~\cite{cdgm}) as well.

\begin{eqnarray*}
\begin{array}{c||c|c}
    \multicolumn{3}{c}{\rm Table~ 2:\ Matrix \ Element\ Results} 
\\ \hline
{\rm Method} & {\cal M}_7 & {\cal M}_8 \\ \hline
{\rm RWM} & 0.16 \pm 0.10 & 2.22 \pm 0.67 \\ \hline
{\rm FESR (ALEPH)} &      & 1.40 \pm 0.28 \\
{\rm FESR (OPAL)}  &      & 1.68 \pm 0.32 \\ \hline
{\rm HYBRID (ALEPH)} & 0.225 \pm 0.046 & 1.55  \pm 0.52 \\
{\rm HYBRID (OPAL)} & 0.210 \pm 0.044 & 1.66  \pm 0.46 \\ 
\end{array}
\end{eqnarray*}

\subsection{Comments}

The content of Table~2 naturally gives rise to two issues which 
require further discussion: 
(i) whether the RWM, FESR and HYBRID results can be combined into a 
single value, and (ii) whether the ALEPH and OPAL values can 
be combined into a single value.  We consider each in turn.

The FESR method described at length in this paper utilizes a 
rather different evaluation procedure from the RWM of 
Ref.~\cite{cdgm}.  The HYBRID method is at first glance 
a consistency check between RWM and FESR which uses the FESR 
estimations of $H_{1,8}$ ({\it i.e.} higher-dimension effects) in 
evaluating the RWM sum rules at the low scale $\mu = 2$~GeV.  
However, it should be understood that, in the low-scale RWM 
evaluation of the integrals $I_1$ and $I_8$, about $60\%$ of the full 
contribution comes from the input values of the chiral constraint 
integrals, and only about $40\%$ from the integrals over the 
$\tau$ data.  Thus the pFESR and hybrid evaluations are 
to a significant extent independent, and the agreement between the 
two represents a highly nontrivial mutual check.  
We note that since the error bars for the 
FESR results are rather smaller than those for RWM, they 
would dominate any averaging procedure.  As a practical matter, 
we choose to simply point out that the FESR approach yields 
our best determination of  ${\cal M}_{8}$, while the 
HYBRID approach leads to our best determination of ${\cal M}_{7}$, 
and leave it at that.

Let us hereafter accept the FESR determination as our `official' 
result.  We discuss next the procedure for combining the ALEPH and OPAL
evaluations of $a_6$ and $a_8$, beginning with the second uncertainties 
in Eqs.~(22),(24). Since these arise in both ALEPH and OPAL analyses
from the same sources ($S_{\rm EW}$, $B_e$, $B_\pi$, $V_{ud}$; OPE
coefficients $a_4$ and $b_6$), they are common to both analyses and
cannot be reduced by averaging. For definiteness, we adopt for this 
common uncertainty the midpoint between the ALEPH and OPAL values. 
As for the first uncertainties in Eqs.~(22),(24), they could be
treated as ordinary independent errors if the ALEPH and OPAL
covariance matrices were fully independent.  However, the covariance
matrices have a common component since the two experiments share
uncertainties due to common normalization input ({\em i.e.}, OPAL used
the 96/97 PDG results for the tau branching ratios and these were
dominated by ALEPH measurements). We do not have the detailed 
knowledge to determine exactly the degree of
correlation induced in the output of our ALEPH and OPAL analyses.
Therefore, we have performed the averaging assuming a generic
correlation coefficient $c$ between the ALEPH and OPAL results. The
following combined results correspond to the conservative value $c=0.5$:
\bea
a_6 &=&  (-48.1 \pm 5.8  \pm 3.1) \cdot 10^{-4} \, \mbox{GeV}^6 \\
a_8 &=&  (-44.3 \pm 26.6 \pm 12.8) \cdot 10^{-4} \, \mbox{GeV}^8 \ \ ,
\eea
Dependence on the correlation parameter $c$ is modest ({\it e.g.} 
for $c=0$ we obtain $a_6 = 
(-48.8 \pm 4.7 \pm 3.1) \cdot 10^{-4} \, \mbox{GeV}^6$ and 
$a_8 =  (-40.9 \pm 21.8 \pm 12.8) \cdot 10^{-4} \, \mbox{GeV}^8$). 
Combining the above uncertainties in quadrature yields 
\be
a_6 =  (-48.1 \pm 6.6) \cdot 10^{-4} \, \mbox{GeV}^6 \ , \qquad 
a_8 =  (-44.3 \pm 29.5) \cdot 10^{-4} \, \mbox{GeV}^8 \ \ .
\ee
Our final combined numbers for ${\cal M}_{7,8}$ in 
the chiral limit are:
\be
{\cal M}_{7}^{\rm chiral}  =   0.22  \pm 0.05 \ , \qquad 
{\cal M}_{8}^{\rm chiral}  =   1.50 \pm 0.27  \ \ . 
\ee

\subsection{Chiral Corrections}\label{sect:chcorr}
The above numbers represent our determination of the 
electroweak penguin matrix elements in the chiral limit, 
where  a data-driven evaluation has been possible. 
However, to make contact with phenomenology, an estimate 
of the chiral corrections is mandatory. This issue has been 
studied by two of us in Ref.~\cite{cg} within NLO Chiral 
Perturbation Theory.  At this order there are two contributions 
to the chiral corrections: one is given by the chiral loops and 
comes with no uncertainty, while the other is due to local couplings 
in the effective theory, not known accurately at present. 
A conservative estimate of these couplings, based on 
naive dimensional analysis (and supported by 
an explicit calculation in leading $1/N_c$~\cite{pps}), gives 
\beq
{\cal M}^{\rm physical}_{7,8} = {\cal M}^{\rm chiral}_{7,8}
 \times F_{\chi} \ \ , 
\eeq
where $F_{\chi}$ represents the dispersive and absorptive corrections 
(generated at NLO) to the matrix elements,~\cite{cg} 
\beq
F_{\chi} = (0.7 \pm 0.2) - i  \, 0.21 \ \ .
\label{fchi}
\eeq
That the determination of $F_\chi$ is less firm than that of 
${\cal M}^{\rm chiral}_{7,8}$ is reflected in its relatively 
larger uncertainty.
Adopting the result of Eq.~(\ref{fchi}), we are led to quote 
\bea
|{\cal M}^{\rm physical}_{7}| &=&  0.16 \pm  0.035_{\chi - {\rm lim}}  
\pm  0.044_{\chi - {\rm corr}} =   0.16 \pm 0.06  \ ,  \\ 
|{\cal M}^{\rm physical}_{8}| &=&  1.10 \pm  0.20_{\chi - {\rm lim}}  
\pm  0.30_{\chi - {\rm corr}}  = 1.10 \pm 0.36  \ .
\eea
We shall use ${\cal M}^{\rm physical}_{8}$ to estimate 
the electroweak penguin contribution to $\epsilon ' / \epsilon $. 

\subsection{Comparison with Other Analyses}

We observe that other analyses of the V-A correlator exist in the
literature~\cite{ioffe,dghs,narison,ppr} which give generally 
different values of the OPE coefficients from those found here 
(especially for the $a_d$ with $d>6$).  The relation between our 
analysis and these others will be discussed in a 
companion paper~\cite{cdgm6}.
 
Here we focus on the comparison with other determinations of ${\cal
M}_{7,8}$ in the chiral limit, based on alternate non-perturbative
methods.  The analytic methods of Refs.~\cite{narison,lund,knecht} 
work directly in the chiral limit, while the lattice results  
refer to chiral limit extrapolations, explicitly reported in
Refs.~\cite{rbc,cppacs,spqr}.  When necessary, we have converted the
lattice results to the $\overline{MS}$-NDR scheme at $\mu=2$ GeV.
Moreover, lattice results are obtained in the quenched approximation
and the quoted error is only statistical.  Finally, let us recall that
other determinations of electroweak matrix elements can be found in
Refs.~\cite{others}.

\begin{center}
\begin{tabular}{c||c|c}
 &  $ {\cal M}_{7} $    &  ${\cal M}_{8} $  \\
\hline 
 This work  &   0.22 $\pm$ 0.05 &  1.50 $\pm$ 0.27  \\
\hline\hline 
Bijnens et al. \cite{lund} &   0.24 $\pm$ 0.03 & 1.2 $\pm$ 0.7 \\
\hline 
 Knecht et al. \cite{knecht} & 0.11 $\pm$ 0.03 & 2.34 $\pm$ 0.73 \\
\hline 
 Narison \cite{narison} & 0.21 $\pm$ 0.05  & 1.4 $\pm$ 0.35 \\
\hline 
\hline
 RBC (DWF) \cite{rbc} & 0.28 $\pm$ 0.04 & 1.1 $\pm$ 0.2  \\
\hline 
CP-PACS (DWF) \cite{cppacs} & 0.24 $\pm$ 0.03 & 1.0 $\pm$ 0.2  \\
\hline 
 SPQ$_{\rm cd}$R (Wilson) 
\cite{spqr}  
& 0.24 $\pm$ 0.02 & 1.05 $\pm$ 0.10 \\
\hline 
 Vacuum Saturation  & 0.32  & 0.94 
\end{tabular}
\end{center}

Our value for ${\cal M}_8$ is slightly higher than the lattice 
determinations.  However, the agreement is better than previous 
work had indicated. The present lattice results are 
larger than earlier lattice estimates while our new result is about
one standard deviation below our previous central value.  Given that
the lattice results are in the quenched limit, the present level of
agreement is satisfactory.  Lattice estimates of ${\cal M}_7$ are also in
agreement with our results.  Finally, the results of
Refs.~\cite{narison},\cite{lund},\cite{knecht} also seem in 
reasonable agreement with ours, with the possible exception of the 
smaller ${\cal M}_7$ result of Ref.~\cite{knecht}.

\section{Conclusion}\label{sect:concl}
This paper has described an improved chiral determination of the
electroweak penguin
contribution to $\epsilon'/\epsilon$. Let us summarize the nature of the
improvement
over our previous result. Recall that our determinations are not
calculations in the
usual sense of constructing an approximation to QCD which is then used
to calculate the operator matrix element. 
Rather, our methods start from the observation that the matrix elements
of the electroweak penguin operators ${\cal Q}_{7,8}$ become related 
in the chiral limit to the vacuum matrix elements of two other
operators, and that these vacuum matrix elements can be determined 
from existing experimental data.  
Our previous determinations and those of the present paper
are based on different approaches to extracting these vacuum
matrix elements. Much of our work has been devoted to minimizing 
the impact on our final uncertainties of both the absence of spectral 
data above $s=m_\tau^2$ and the presence of large experimental 
errors near the upper end of the kinematically accessible 
range. A detailed understanding of these uncertainties has also 
been obtained.

Our earlier chiral determination of 
$\left[\epsilon'/\epsilon\right]_{\rm EWP}$ 
({\it cf.} Eq.~(\ref{old})) was based on the use of other rigorous 
chiral sum rules in order to minimize the effect where the data 
were poor or non-existent. This is a very direct approach, and 
we found that it carried a 32\% uncertainty. The error bar was 
partially due to the residual experimental uncertainty from the tau
decay data. However the error bar also had a large component due to 
uncertainties in the inputs to the other chiral sum rules which were used 
as constraints, the largest of which involved the pion's 
electromagnetic mass difference in the chiral limit.  For example, 
at $\mu =4$ GeV (where it is safe to neglect higher dimensional 
contributions) the result was a determination of ${\cal M}_8$ with
$32\%$ uncertainty. Roughly $80\%$ of the error on the dispersive
integral $I_8(4\ {\rm GeV})$ is due to uncertainties in the inputs to 
the chiral sum rule constraints, the dominant contribution being
the above-mentioned uncertainty in the chiral limit value of the 
pion electromagnetic mass splitting.

The present evaluation, based on the FESR method, does not use the 
chiral sum rules and hence does not share the same uncertainties. 
Rather, the FESR error bars arise primarily from uncertainties in the 
tau decay data.  We use doubly pinched 
weights in the FESR integrals both to minimize the influence of 
these data uncertainties and also to suppress OPE contributions from 
the vicinity of the timelike real axis.  Presumably, the latter effect 
serves to improve the reliability of the 
OPE representation for $\Delta\Pi$.  To perform 
a highly nontrivial check of 
this presumption, we extract each OPE coefficient $a_d$ 
with more than one weight.\footnote{The motivation behind this 
strategy is that, if using the OPE were indeed dangerous ({\it e.g.} 
the integrated OPE therefore gives a poor representation of 
the data), then one should expose this 
flaw by using both weights simultaneously.} 
In all cases, excellent 
consistency is found.  An instructive 
demonstration is provided by Figs.~3,4, which involve 
the determination of $a_6$ and $a_8$ using the weights $w_1$ and $w_2$.
By design, $w_1$ and $w_2$ have very different profiles and 
thus probe the spectral function in very different ways. 
The spectral integrals for both the $w_{1,2}$-weighted FESR's 
turn out to scale very closely with $1/s_0^2$ (each case implying 
that the $d=8$ contribution is small) and the consistency between 
the $a_6$ values thus extracted turns out to be excellent. 
Our procedure has passed a number of additional tests, as will 
be described elsewhere.~\cite{cdgm6}  
Therefore, once one makes sure that the systematics of the FESR
approach are under control, the FESR analysis leads to a more precise
determination of ${\cal M}_8$, because it exploits in an optimal 
manner the existing database and (unlike the RWM) does not 
rely on other input.

When we translate the measurement of the OPE coefficient into the
electroweak penguin contribution to $\epsilon'$, we find
\beq
\left[\epsilon'/\epsilon\right]_{\rm EWP, \chi-{\rm lim}} = 
\left( - 15.0 \pm 2.7 \right) \cdot 10^{-4} \ \ , 
\label{new}
\eeq
in the chiral limit. This is consistent with our previous determination,
but has about half (at $18\%$) the uncertainty. When we include the 
chiral corrections as described in Sect.~\ref{sect:chcorr}, we find 
\beq
\left[\epsilon'/\epsilon\right]_{\rm EWP, \, {\rm phys}} = 
\left( - 11.0 \pm 3.6 \right) \cdot 10^{-4} \ \ .
\label{new2}
\eeq
The larger error (about $32\%$) cited for the `physical' result 
reflects uncertainties in our estimate for the chiral corrections. 
We note that for the Standard Model to successfully 
describe $\epsilon'$ will require a 
rather large and positive effect from the standard gluonic penguin
operator ${\cal Q}_6$. Our approach does not provide a determination 
of this matrix element.

In addition to describing the electroweak penguin contribution to
$\epsilon'$, our work is useful in other contexts. We have a firm 
determination of the ${\cal Q}_8$ nonleptonic matrix element in 
the chiral limit as well as that for ${\cal Q}_7$, and 
there are few such examples that are experimentally known. The results
are useful for comparison with models, some of which are included in 
the comparisons Table in Sect.~IV-C, 
or for the testing of lattice methods. Our methods 
may also be adapted rather directly to lattice techniques, because 
the correlation functions that we work with are readily measured with 
lattice data. We have started work to make this connection firmer,
and will report on the results in the future.

\acknowledgements
The work of J.D. and E.G. was supported in part by the National
Science Foundation under Grant PHY-9801875.
The work of V.C. was supported in part by MCYT, Spain (Grant No.
FPA-2001-3031) and by ERDF funds from the European Commission.  
K.M. would like to acknowledge the ongoing
support of the Natural Sciences and Engineering Research
Council of Canada. We are happy to acknowledge useful input from 
S. Menke, M. Papinutto, S. Peris, A. Pich, J. Prades and M. Roney.

\eject

\end{document}